# Deep Atmosphere Composition, Structure, Origin, and Exploration, with Particular Focus on Critical *in situ* Science at the Icy Giants

Sushil K. Atreya[1], Mark H. Hofstadter[2], Joong Hyun In[3], Olivier Mousis[4], Kim Reh[2], and Michael H. Wong[5]

[1]University of Michigan, Climate and Space Sciences and Engineering, 2455 Hayward Street, Ann Arbor, MI 48109-2143, USA (atreya@umich.edu)
[2]Caltech Jet Propulsion Laboratory, 4800 Oak Grove Drive, Pasadena, CA 91109, USA
[3]School of Marine and Atmospheric Sciences, Stony Brook University, Stony Brook, NY 11794, USA
[4]Laboratoire d'Astrophysique de Marseille, Aix Marseille Université, CNRS, CNES, Marseille, France
[5]Department of Astronomy, University of California, Berkeley, CA 94720, USA

## Abstract

Core accretion is the conventional model for the formation of the gas giant planets. The model may also apply to the icy giant planets, Uranus and Neptune, except that it may take upward of 50 Myr for them to form at their present orbital distances, which is beyond the maximum 5 Myr lifetime of the solar nebula. A plausible alternative is formation in the region of the gas giants, followed by migration to their present locations at 20 and 30 AU. Another alternative is the gravitational instability model, which is much faster and does not require the formation of a core first. In either scenario, heavy elements (mass > helium) provide the critical observational constraints. Additionally, helium and neon abundances in the observable troposphere are indicators of the interior processes in the megabar region. We investigate the atmospheric regions most suitable for accessing the above elements. Volatiles containing some of the elements (C, N, S, O) undergo condensation on the icy giants. On the other hand, noble gases (He, Ne, Ar, Kr, Xe), which are chemically inert, non-condensible, and uniform all over the planet, can provide the best constraints to the formation and migration models of Uranus and Neptune. Only entry probes are capable of measuring the key elements and isotopic ratios. They are accessible at 5-10 bars, except for the condensibles. Data from an orbiter on gravity, magnetic field, upper atmospheric composition and the maps of ammonia and water with depth would be a valuable complement to in situ measurements.

## 1 Introduction

The giant planets are key to understanding how the solar system formed, and, by extension, how other solar systems around sun-like stars form. Due to their large mass and substantial magnetic fields, loss of volatiles, even the lightest element, hydrogen, from the giant planets is insignificant since the time of their formation 4.6 billion years ago. Thus, the giant planets are arguably the ideal objects to reveal the conditions of the primordial solar nebula from which all solar system bodies formed. Whereas Jupiter and Saturn are largely gaseous, with solid material comprising only about 3-5% of their planetary mass, Uranus and Neptune are made up of mostly icy/rocky material, which could be up to 90% of their mass. Thus, the icy giant planets are in a class of their own, distinct from the gas giant planets, yet very much a part of the giant planet family. While multiple spacecraft have explored Jupiter and Saturn extensively since the early 1970's, much of what we know about the properties of the icy giant planets is the result of brief flybys of Uranus and Neptune by the Voyager 2 spacecraft, respectively in 1986 and 1989, and observations from 1 AU. However, the data to date on their atmospheric composition, structure, dynamics, magnetic fields, interiors, rings and satellites, etc. is still quite limited in scope. A comprehensive study of the icy giant planets is thus essential to understand their formation and migration scenarios as well as their role in the formation of the outer solar system. Uranus and Neptune also serve as the only solar system analogs for the most common type of nearly 4000 confirmed exoplanets to date, about 40% of which are mini-Neptune to Neptune size objects (Figure 1). The focus of this paper is mainly on the formation and the compositional data necessary to constrain them. We will first discuss briefly the conventional formation models, followed by a discussion of key sets of data and an outline of possible scenarios for a mission to the icy giant planets.

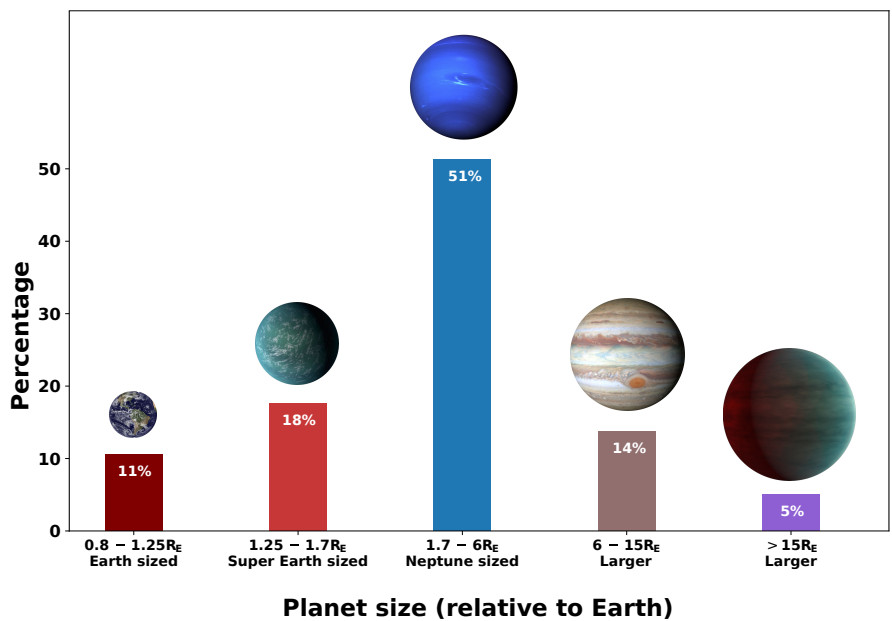

**Figure 1.** Mini-Neptune to Neptune-size exoplanets (1.7–6 $R_E$) make up about one-half of nearly 3100 exoplanets with known radii out of a total of about 4000 confirmed exoplanets to date (2020 January). Percentages are rounded off to the nearest whole number; the actual values are 10.53%, 17.61%, 51.28%, 13.77% 5.08% and 1.73%, respectively, for 0.8-1.25 $R_E$, 1.25-1.7$R_E$, 1.7-6$R_E$, 6-15$R_E$, >15$R_E$ and <0.8$R_E$ (not shown) size ranges. [Illustration based on listings in *NASA Exoplanet Archive*: https://exoplanetarchive.ipac.caltech.edu, and *Extrasolar Planets Encyclopedia*: http://exoplanet.eu]

## 2 Formation Scenarios

The presence of Uranus and Neptune in our solar system raises the question of their formation in the framework of the standard theories of planetary formation. Both current formation models, namely the core accretion and the gravitational instability models, have been used to explain the formation and properties of the two planets. Formation by core accretion is a slow and steady process, whereas formation by gravitational stability is relatively fast.

In the core accretion model, the formation of a giant planet starts with non gravitational collisions between micron size dust grains (of metals, ices, possibly refractories, and trapped volatiles) resulting in relatively large planetesimals, followed by core growth, concurrent accretion of solids and gas onto the core, and finally by the rapid accretion of a massive gaseous envelope once the core reached a critical mass of 10-15 Earth Masses, $M_E$, (Mizuno 1980; Pollack et al. 1996; Hubickyj et al. 2005). In this scenario, had Uranus and Neptune formed at their current orbits of 20 and 30 AU, they would require formation timescales on the order of 50 Myr due to lower disk density at those distances (Pollack et al. 1996), i.e. on a time scale that far exceeds the 1-5 Myr lifetime of the protosolar nebula (PSN). Planetary migration, which is commonplace in exoplanets, has therefore been invoked during the growth of the two icy giants planets in the neighborhood of Jupiter and Saturn (Trilling et al. 1998; Edgar 2007; Alexander and Armitage 2009; Helled and Bodenheimer 2014). On the other hand, recent formation models suggesting the accretion of the icy giants from pebbles may also overcome the timescale problem (Lambrechts and Johansen 2012; Johansen and Lambrechts 2017). In all scenarios, the metallicity of the icy giants strongly depends on the composition of the accreted planetesimals or pebbles. The high elemental abundances observed in the envelopes would result either from the devolatilization of the accreted solids or from the dilution of the cores, or both. In contrast with the core accretion model, giant planets form rather rapidly in tens of thousand years directly from gas as a result of gravitational instabilities in a disk (Boss, 1997; Mayer et al., 2002). In this case, the growth of disk perturbations leads to density enhancements or clumps in disk regions, where self-gravity becomes as important as, or exceeds the stabilizing effects of pressure and shear. It has been proposed that icy giants could consist of remnants of gas giants that formed from disk instability, and whose cores would have formed from the settling of dust grains in the envelopes prior to their photoevaporation by a nearby OB star (Boss et al., 2002). Planets formed in this manner could also present envelopes with enhanced metallicities compared to that of their host star, considering the possibility of later addition of material. Though gravitational instability model might seem to be consistent with global properties of Uranus and Neptune's, it has at least two major shortcomings: (a) it requires a stochastic event to occur, and (b) the presence of natural and captured satellites around these planets suggest that planetesimal and/or pebbles were already present during planet formation, thus reinforcing the idea of the core accretion scenario. Also, if one assumes that the four giant planets formed into a compact system, then Jupiter and Saturn would also have suffered from photoevaporation, which is not the case, given their much lower enrichments in volatiles, compared to Uranus and Neptune (provided that their C/H enrichment is proxy for the other heavy elements).

The core accretion model is further strengthened by several observations, including heavy element enrichment at Jupiter, and where available, the other giant planets, first solids (millimeter size chondrules and calcium aluminum inclusions) at the very beginning of the solar system, and greater frequency of exoplanets around higher metallicity stars (see Atreya et al. 2019a and references therein for additional details).

Formation of a massive core is central to the core accretion model. The core is made from "heavy elements" ($>^4He$), which were presumably released to the envelope during accretionary heating phase, thus forming the atmosphere, together with the most volatile of gases, hydrogen, helium and neon, which were gravitationally captured from the surrounding protoplanetary nebula when the core reached a critical mass. The heavy elements are thus key constraints to the formation models. Though much is known about Jupiter's heavy elements, little information presently exists for the icy giant planets, since it requires in situ measurements of bulk composition, which are presently lacking for the icy giants. In the following section we discuss the presently known information and expectations for their elemental and isotopic abundances, followed by required measurements.

## 3 Bulk Composition, and Vertical Variation

Constituents in the upper atmosphere and the ionosphere are subject to meteorological, dynamical, circulation and chemical processes, which govern their distribution. The region of the atmosphere below which the relative proportion of a given species remains constant is where the species elemental abundance can be determined. In the atmosphere of the icy giants, very few molecules have been detected so far, and a vast majority of them in their stratospheres. Figure 2 illustrates this point for Neptune, but is nearly equally valid for Uranus. Remote sensing in the UV allowed the determination of certain hydrocarbons in the upper atmosphere (Broadfoot et al. 1986, 1989). IR detected certain hydrocarbons in the stratosphere and $H_3+$ higher (Fletcher et al. 2010; Meadows et al. 2008, Orton et al. 2014, Trafton et al. 1993, Stallard et al. 2014). Radio occultations from spacecraft made measurements of the electron concentration higher up (inset; Tyler et al. 1986, 1989) and of temperature in the lower stratosphere (Lindal et al. 1992, 1987), whereas the ground-based VLA data in the microwave are able to sense deeper in the troposphere (Gulkis et al. 1978). This illustration also shows possible existence of several cloud layers of multiple species at different depths and a water cloud/ocean at 10's – 100's of kilobars. Thus the well-mixed atmosphere for measuring the bulk composition, hence the elemental abundances, is most likely quite deep for the condensible species. In the following subsections, we first discuss the condensible species, followed by the noble gases.

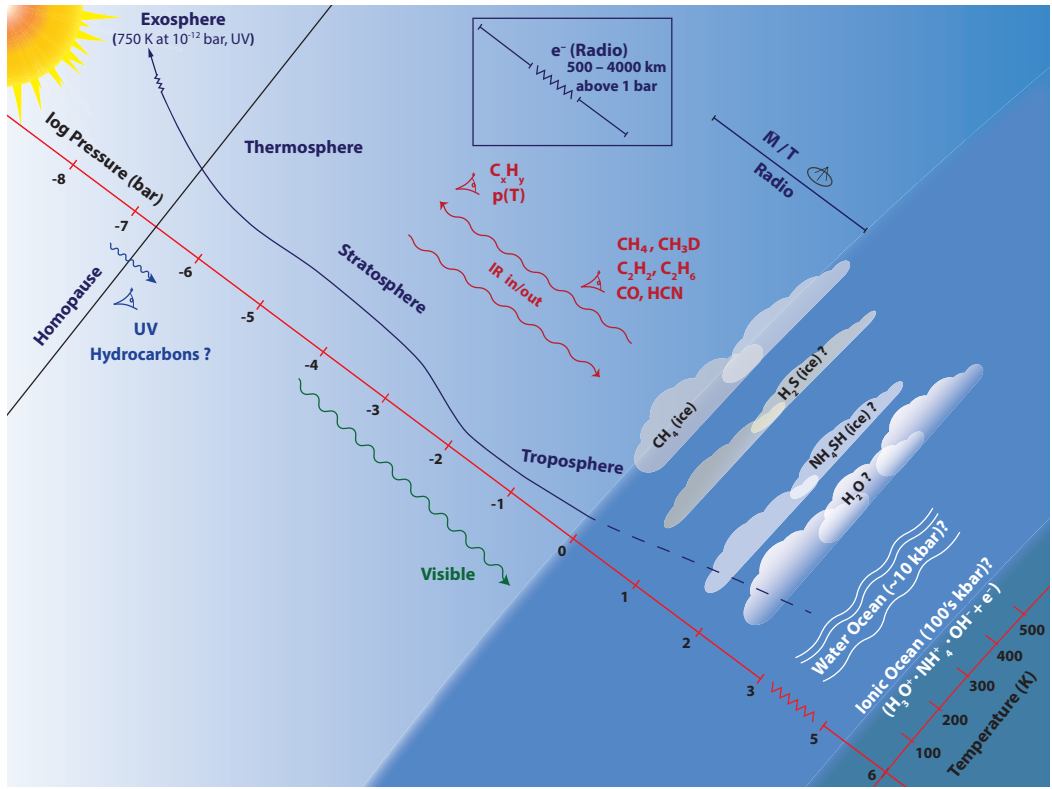

**Figure 2.** Illustration of the regions of Neptune's atmosphere that can be explored using different parts of the solar spectrum and the type of information obtained. For example, UV is useful down to ~10 microbar, whereas radio occultations are good for the ionosphere (inset) and again in the troposphere between ~1-1000 millibar. Only the topmost cloud layer has been inferred from the radio occultation observations done on Voyager. The composition of this cloud and others shown in the figure are based on thermochemical models. Maarten Roos-Serote helped with an earlier version of the graphic. [Fig. 1 in Atreya et al., Acta Astronautica, 162, 266, 2019b]

## 3.1 Condensibles: Methane, Ammonia, Hydrogen Sulfide, and the Elemental Abundances of C, N And S

Current elemental abundances of all four giant planets are shown in Figure 3 (Atreya et al. 2019a). The actual values together with their ratios to protosolar abundances are given in Table 1. The bulk of a giant planet's carbon is sequestered in methane ($CH_4$). In Uranus and Neptune, the only elemental abundance determined to date is that of C from ground-based observations of $CH_4$. The C/H ratio is found to be 80±20× solar in both icy giants, although Figure 4 shows that deriving the C/H elemental ratio from atmospheric $CH_4$ mixing ratio retrievals involves assumptions about the amount of H tied up in volatiles such as $H_2O$ and $NH_3$. Remote-sensing retrievals of $CH_4$ atmospheric mixing ratios are complicated by latitudinal gradients presumably established by meridional circulation (e.g., Karkoschka and Tomasko 2011, Tollefson et al. 2019, Sromovsky et al. 2019), and it is unknown whether the level of well-mixed $CH_4$ may be displaced downward by multiple scale heights due to dynamical effects, as Juno found for ammonia at Jupiter (Bolton et al. 2017; Li et al. 2017).

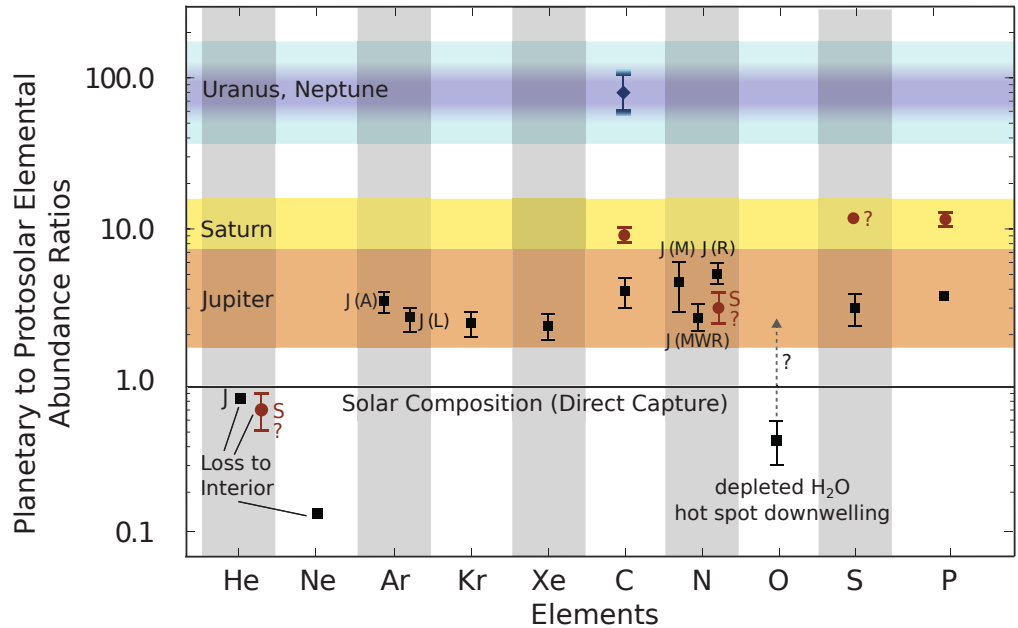

**Figure 3.** Elemental abundance ratios in the atmospheres of Jupiter, Saturn, Uranus and Neptune relative to the protosolar values. "N" in Jupiter represents values from ammonia ($NH_3$) abundance measurements made by the Galileo probe mass spectrometer [J(M)] and the attenuation of probe radio signal [J(R)] as well as the Juno microwave spectrometer [J(MWR)]. The Jovian "Ar" value from the Galileo Probe measurements is shown as ratio to the protosolar Ar/H value of Asplund et al. [J(A)] and Lodders et al. [J(L)] since the solar argon values of the two sets of authors differ considerably. Saturn's He and N are labeled S. N/H of Saturn is a lower limit, and S/H is highly questionable. Only C/H is determined for Uranus and Neptune from ground-based $CH_4$ to ~1bar,level, but could be greater in the deep atmosphere (see Atreya et al. (2019a) and references therein for other details). [This version of the figure was adapted from Fig. 2.1 of Atreya et al. (2019a), with permission from Cambridge University Press, PLSclear Ref No: 18694].

For all practical purposes, ammonia ($NH_3$) is the bulk reservoir of nitrogen in all giant planets. The earliest radio observations of Uranus showed that, unlike Jupiter and Saturn, its atmospheric composition was far from solar (Gulkis et al. 1978). Subsequent observations a decade later confirmed that and found even larger depletion factors in some places on Uranus and Neptune (de Pater et al. 1989, 1991). Gulkis et al. (1978) found in particular that $NH_3$, the dominant opacity source at Jupiter and Saturn, must be depleted relative to its solar abundance by a factor of 100 at temperatures below ~250 K (pressures < 40 bars). This was the best way to explain the very bright radio emissions seen at wavelengths from ~1 to 6 cm, as $NH_3$ is a powerful microwave absorber and even modest amounts of it would shield the hotter regions of the atmosphere from view. Even lower values were reported in later publications. The observed depletion of $NH_3$ abundance at these temperatures is surprising because the bulk of the planet is expected to be enriched in $NH_3$ relative to solar, and contrary to the predictions of equilibrium cloud condensation models, an example of which is shown in Figure 5. This figure is based on temperatures measured on Voyager using the radio occultation technique at Uranus (Lindal et al. 1987) and their $CH_4$ mole fraction of ~45× solar (the current $CH_4$ values are higher, Table 1). The elemental ratios of N, S and O are arbitrarily taken as 1× solar, only for illustration purposes. The $NH_3$ cloud in this case forms at ~10 bars. A $CH_4$ cloud also forms in 1-1.4 bar range (Figure 5), consistent with the Voyager finding of a cloud layer in this region. [Other cloud layers in Figure 5 are an $NH_4SH$ ice and the ice and droplet clouds of $H_2O$ that result from assumed atmospheric composition.] Below its cloud level, $NH_3$ vapor is expected to be well-mixed, i.e. 1× solar in this model, not depleted by factors of 100 or so measured by the VLA data. As the temperature of Neptune at 1 bar is within only a few degrees of Uranus and the lapse rates are expected to be similar, the result is nearly the same as for both planets. In Jupiter also, depletion of ammonia to several tens of bars below its 0.7 bar condensation level has been measured by the Juno microwave radiometer, but the depletion factor is only about 2 (Bolton et al. 2017, Li et al. 2017).

Table 1. Elemental abundances in the Sun, Jupiter, Saturn, Uranus and Neptune, and planetary to protosolar ratios

| Elements | Sun-Protosolar[a,b] | Jupiter[c] | Saturn[c] | Jupiter/Protosolar | Saturn/Protosolar | Uranus/Protosolar | Neptune/Protosolar |
|---|---|---|---|---|---|---|---|
| He/H | $9.55×10^{-2}$ | $7.85±0.16×10^{-2}$ | $6.75±1.25×10^{-2(c,d)}$ $6.20±1.25×10^{-2(e)}$ | $0.82±0.02$ | $0.71±0.13$ $0.65±0.13$ | $0.94±0.16^{(f)}$ | $1.26±0.21^{(f)}$ $0.94±0.16^{(f)}$ |
| Ne/H | $9.33×10^{-5}$ | $1.24±0.014×10^{-5}$ | Not available (NA) | $0.13±0.001$ | NA | NA | NA |
| Ar/H | $2.75×10^{-6}$ | $9.10±1.80×10^{-6}$ | NA | $3.31±0.66$ | NA | NA | NA |
| Kr/H | $1.95×10^{-9}$ | $4.65±0.85×10^{-9}$ | NA | $2.38±0.44$ | NA | NA | NA |
| Xe/H | $1.91×10^{-10}$ | $4.45±0.85×10^{-10}$ | NA | $2.34±0.45$ | NA | NA | NA |
| C/H | $2.95×10^{-4}$ | $1.19±0.29×10^{-3}$ | $2.65±0.10×10^{-3}$ | $4.02±0.98$ | $8.98±0.34$ | $80±20^{(g)}$? | $80±20^{(h)}$? |
| N/H | $7.41×10^{-5}$ | $3.32±1.27×10^{-4(i)}$ $2.03±0.46×10^{-4(j)}$ | $2.27±0.57×10^{-4}$ ($f_{NH_3}=4±1×10^{-4}$) | $4.48±1.71$ $2.70±0.60$ | $≥3.06±0.77$ | $0.0–0.001^{(k)}$? | $0.01–0.001^{(k)}$? |
| O/H | $5.37×10^{-4}$ | $2.45±0.80×10^{-4}$ (5-μm hotspot) | NA | $0.46±0.15$ | NA | NA | NA |
| S/H | $1.45×10^{-5}$ | $4.45±1.05×10^{-5}$ | $1.88×10^{-4}$ (?) | $3.08±0.73$ | $13.01$ (?) | $>(\sim 0.4–1.0)^{(l)}$? | $>(\sim 0.1–0.4)^{(m)}$? |
| P/H | $2.82×10^{-7}$ | $1.08±0.06×10^{-6}$ | $3.64±0.24×10^{-6}$ | $3.83±0.21$ | $12.91±0.85$ | NA | NA |

NA: Not available

[a]Protosolar values based on the solar photospheric values of Asplund et al. (2009, table 1).
[b]Protosolar metal abundances relative to hydrogen can be obtained from the present day photospheric values (table 1 of Asplund et al., 2009), increased by +0.04 dex, i.e. ~11%, with an uncertainty of ±0.01 dex; the effect of diffusion on He is very slightly larger: +0.05 dex (±0.01).
[c]Atreya et al. (2019a) and references therein. Saturn's N/H is based on observations down to only 3 bars, which does not preclude higher values in the deep well-mixed atmosphere. Additional details on helium abundance are given separately in Table 2 and plotted in Figure 8.
[d]Conrath and Gautier (2000); [e]Koskinen and Guerlet (2018); [f]Gautier et al. (1995). Two values are given for Neptune, one without $N_2$ in the atmosphere (larger He/H) and the other including $N_2$ in order to explain presence of HCN (see text); [g]Sromovsky et al. (2011); E. Karkoschka and K. Baines, personal communication (2015); [h]Karkoschka and Tomasko (2011); [g,h]C/H based on $CH_4$ measurements to ~ 1 bar level, but could be greater in the deep well-mixed atmosphere depending on atmospheric dynamics; [i]Wong et al. (2004, Galileo); [j]Bolton et al. (2017, Juno), Li et al. (2017, Juno); [k]May not be representative of deep atmosphere (see text); [l]Irwin et al. (2018), [m]Irwin et al. (2019); [l,m]*Lower limit* below an $H_2S$ cloud, based on the detection of $H_2S$ gas in the 1.2–3 bar region above the cloud, thus S/H is not necessarily representative of the deep well-mixed atmosphere value (see text).

The VLA data indicated that higher opacity was required in the shallow troposphere than that due to observed amounts of ammonia. Improved measurements from the ground at wavelengths from the far infrared to the radio (Orton et al. 2014), and improved ab initio calculations of collision-broadened $H_2$ (Orton et al. 2007) also support the idea of more opacity near the ~1 bar region and that a species more volatile than $NH_3$ was needed to provide it. That species was predicted to be $H_2S$ gas with an $H_2S$ ice cloud in the ~3 bar region (de Pater et al. 1989, 1991). Presence of $H_2S$ at these altitudes may plausibly be attributed to an unexpectedly large depletion in $NH_3$ (Atreya and Wong 2005, Atreya et al. 2019b). This is because $NH_3$ normally serves as a sink for $H_2S$ mole for mole, culminating in the formation of a thermochemical cloud of ammonium hydrosulfide ($NH_4SH$) or ammonium sulfide (($NH_4)_2S$) ice at appropriate temperatures in a solar composition atmosphere (Lewis and Prinn 1970). With $NH_3$ severely depleted in the icy giants, $H_2S$ vapor could survive to the upper troposphere, but that requires the S/N ratio to be ≥5× solar (de Pater et al.. 1989, 1991; Irwin et al 2018). The $H_2S$ vapor would condense into an $H_2S$ ice cloud at a pressure level that depends on the assumed enhancement of the S/N ratio above solar, as seen in Figure 6. Indeed, with S/N ratio of about 5× solar, an $H_2S$ ice cloud forms in the 2-3 bar region. This calculation is for illustration purposes only to show the likelihood of an $H_2S$ cloud, whose location can be fine-tuned by adjusting any number of parameters including the atmospheric temperature structure, degree of $NH_3$ depletion in the relevant region, distribution of absorbers, etc. Presence of $H_2S$ vapor and cloud in approximately 1-3 bar region on both Uranus and Neptune has now been confirmed from ground-based observations in the infrared (Irwin et al. 2018, 2019, and Table 1). The methane cloud above it forms the topmost cloud layer on Uranus and Neptune.

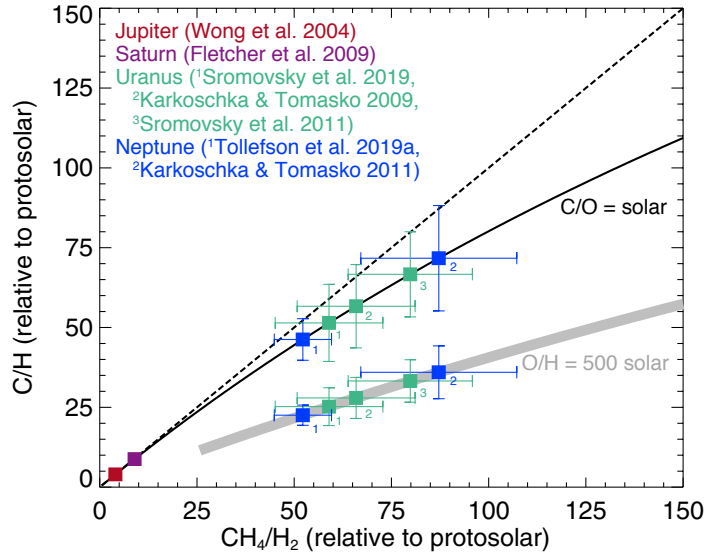

**Figure 4.** The C/H elemental ratio (y-axis) is non-linearly related to measured $CH_4$ hydrogen mixing ratios (x-axis), especially in compositions with high heavy-element abundances. The difference between solid and dashed lines accounts for the amount of hydrogen tied up in the volatile species $CH_4$, $H_2O$, and $NH_3$ throughout the planetary envelope. In the C/O and C/N solar case (solid black line), the 87× solar $CH_4$ mixing ratio for Neptune (Karkoschka and Tomasko 2011) actually corresponds to a 72× solar C/H ratio (a 22% difference). In the 500× solar O/H case (Sec. 4.3.1), so much hydrogen is tied up in $H_2O$ molecules that the observed $CH_4$ mixing ratio would imply only a 36× solar bulk C/H ratio in the envelope.

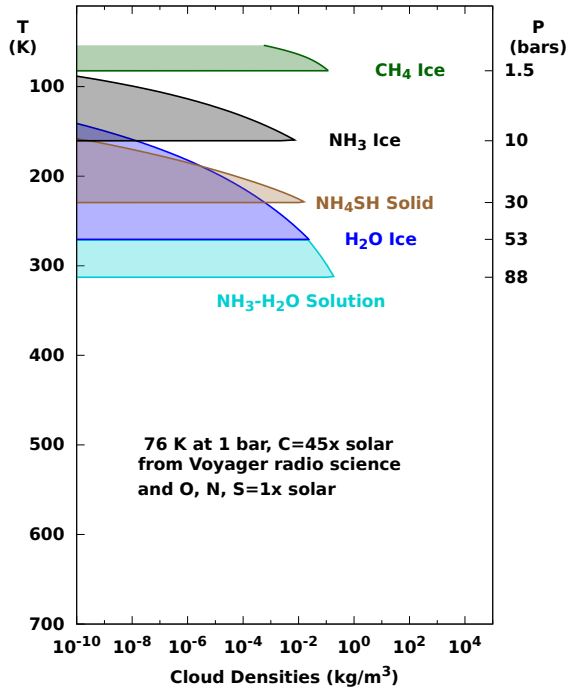

**Figure 5.** A solar-composition cloud structure model, constrained by the p-T and the $CH_4$ mole fraction (C/H = 45× solar) from Voyager [Lindal et al. 1987], would have an $NH_3$-ice cloud layer but no $H_2S$-ice cloud. The results for Neptune are very similar, considering their temperatures are within a few degrees of each other at the 1 bar level. Cloud densities are calculated using the formula from Weidenschilling and Lewis (1973), which Wong et al. (2015) showed corresponds to a very large updraft length scale, equal to the pressure scale height $H(z)$ at each altitude. Actual cloud densities in most Uranus/Neptune situations are likely to be much smaller than shown, but the pressure-levels of cloud bases are unaffected by the choice of updraft length scale.

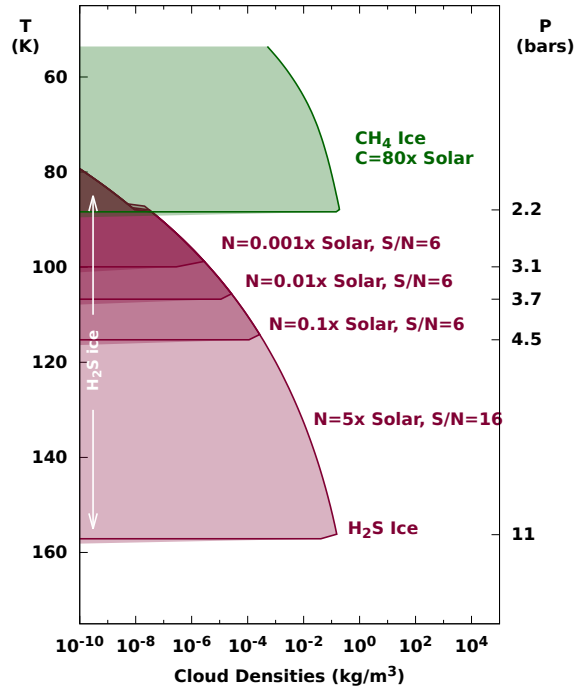

**Figure 6**. As opposed to the solar-composition case of Fig. 4, compositions with supersolar S/N ratios (in this part of the upper troposphere) result in $H_2S$ ice condensation instead. Models consider $NH_3$ of 0.001 to 5 × solar and S/N enhanced by factors of 6 and 16 above solar. $CH_4$ ice continues to remain as the topmost cloud layer, with $H_2S$ ice below at ~3 bar level for S/N=6. See Fig. 5 caption for discussion of cloud densities and cloud base pressure levels.

Although $NH_3$ is found to be depleted in the tropospheres of Uranus and Neptune, it was clear from the longer-wavelength radio data that the depletion of absorber (whether $NH_3$ or not) does not persist throughout the troposphere, as it rapidly becomes more opaque at higher temperatures and pressures (de Pater et al. 1989). The increase in opacity may be associated with a combination of water vapor and water drops (perhaps with $NH_3$ in solution), though an additional absorber (such as increased $NH_3$ vapor or $H_2S$) would improve the fit. The exact abundances and even the identification of some species is subject to many assumptions, but it appears that on Uranus $H_2S$ is modestly enriched relative to solar near 50 bars, while $NH_3$ is much more significantly depleted relative to solar. The S/N ratio in this region is >1, with near solar $H_2S$ in the ~5 to 30 bar region. On Neptune, the trends are the same, though $H_2S$ is more strongly enriched both above and below the $NH_4SH$ cloud. Figure 7 shows a representative VLA observations (Hofstadter et al. 2018). As can be seen, a solar composition atmosphere does not fit the data, but requires high depletion of $NH_3$ in the upper troposphere and $H_2S$ possibly being the main absorber down to ~50 bars. It should be stressed that the $NH_3$ and $H_2S$ abundances even at 50 bars may not be representative of their bulk abundances in the deep well-mixed atmosphere, hence not valid for deriving the true N and S elemental abundances in Uranus and Neptune (Sec. 3.2).

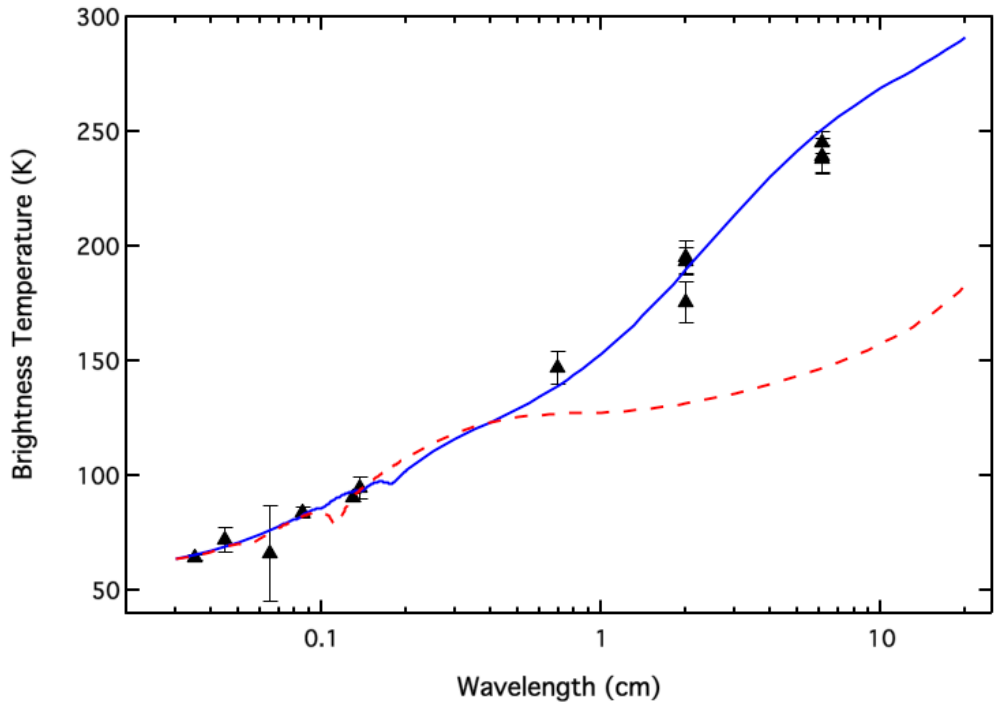

**Figure 7**. A graphic showing a representative VLA spectrum (data points are black triangles), along with a solar abundance model spectrum (red dashed curve) and a best-fit model spectrum (blue solid line).

**3.2 Water: Ionic Ocean and Water Ocean as Potential Sinks of $NH_3$, $H_2S$ and $H_2O$**

The observed depletion of ammonia to tens of bar levels in the icy giant planets is surprising. Predictions based on nearly uniform enrichment of heavy elements measured at Jupiter (Figure 3) and the formation models are that it should be enriched by about the same factor as $CH_4$, i.e. 60-100× solar, not depleted. [Some exceptions may apply to the noble gases whose delivery scenarios may govern their abundances, as discussed later.] So, that begs the question why $NH_3$ and $H_2S$ are so much depleted to great depths. The answer may lie in the distribution and fate of water on Uranus and Neptune. Perhaps, very deep in the interiors of Uranus and Neptune $NH_3$ has high enrichment but some process is sequestering it and possibly also $H_2S$ down there. Atmospheric dynamics alone is unlikely to explain the depletion, considering the high degree of observed depletion. Two potential sinks for $NH_3$ and $H_2S$ are a liquid water ocean at 10 kilobar or deeper and an ionic/superionic ocean at 100's of kilobars.

The liquid water ocean is distinct from an aqueous ammonia cloud predicted by equilibrium thermodynamics models at a kilobar level (Sec. 4.1). As $NH_3$ and $H_2S$ dissolve in water, their abundances would be depleted above such purported water ocean. However, there are two caveats: the degree of loss of $NH_3$ and $H_2S$ in such an ocean and the very likelihood of the ocean's existence. In a solar mix of gases, only 3% of ammonia may be removed by water at 300 K, and it drops at higher temperatures, where a water ocean might form. On the other hand, the solubility may be different at high pressures corresponding to the ocean depth. Preliminary modeling of the likelihood of a liquid water ocean between the cloud tops and the $H_2$-rich deep interior found that Neptune is both too warm and too dry to form such an ocean (Wiktorowicz and Ingersoll 2007), and the same should apply to Uranus as well. Only future laboratory measurements and modeling will tell whether a liquid water ocean actually exists, and, if so, what quantities of $NH_3$ and $H_2S$ it may actually remove.

An ionic ocean provides another means for sequestering $NH_3$ and $H_2S$ in the deep atmospheres of the icy giant planets. Presence of such an ocean was first suggested to explain the relatively large intrinsic magnetic field of Uranus measured by Voyager 2 (Ness et al. 1986), since metallic hydrogen, which drives the internal dynamo at Jupiter and Saturn, is not expected to form in the interiors of cooler, smaller icy giant planets. Subsequently, molecular dynamics calculations and experiments employing Raman spectroscopy in a laser heated diamond anvil cell showed a superionic phase of water forming at temperatures above 2000 K and pressures of 30 GPa (Goldman et al. 2005 and references therein for associated lab experiments). Ionic/superionic water is likely to incorporate ammonia into it (Atreya and Wong 2004, 2005), so that a plausible composition of such an ocean is [$H_3O^+ \cdot NH_4^+ \cdot OH^-$] together with free electrons to maintain charge balance in the plasma. New ab initio calculations also support formation of a stable ammonia-water material at high pressures (Robinson et al. 2018). Such an ocean might take in some hydrogen sulfide also. Thus, an ionic/superionic water ocean is a potential sink for ammonia and possibly $H_2S$ in the interiors of Uranus and Neptune, resulting in their depletion at shallow tropospheric levels sensed by the VLA. Could such an ocean also serve as a sink of water? If water only changed to an ionic phase, it would not mean loss of water, just as hydrogen changing to metallic hydrogen in the interior of Jupiter does not imply loss of hydrogen. However, if water combines with other species such as $NH_3$ and $H_2S$ it could "bind" to the larger compound, which may serve to sequester water itself. The actual depletion of water would depend on the degree to which water is ionized relative to its total reservoir and the extent and efficiency of its binding properties with other compounds. Additional modeling and laboratory studies are needed to test the feasibility of sequestration of volatiles in the purported ionic/superionic ocean.

**3.3 Noble Gases**

Noble gases are particularly important to the models of planetary formation, deep interior processes and the energy balance. Being chemically inert and non-condensible, the noble gases are not subject to variations over the planet unlike other volatiles discussed above. Though not a heavy element, helium has a special role in the giant planets. It may even control the fate of a heavy noble gas, neon, so we divide the discussion of the noble gases into two parts, one on He and Ne and the other on Ar, Kr and Xe.

**3.3.1 Helium, Neon, And The Planetary Interior and Heat Balance**

Helium condenses in the 1-2 megabar region in Jupiter's interior. Neon dissolves in liquid helium (Roulston and Stevenson 1995; Wilson and Militzer 2010). Helium rain drops carry neon with them. As a consequence, the abundance of helium and neon would be depleted in the sensible troposphere. Galileo probe found just that. Helium was only 80% of its expected solar ratio to hydrogen, while neon was only 10% solar (Figure 8, Table 2). Latent heat released upon condensation of helium together with the conversion of potential energy to kinetic energy on differentiation of helium drops from molecular hydrogen gas would contribute to internal heat of the planets. Both Jupiter and Saturn emit nearly twice as much heat as they absorb from the Sun. A good fraction of the internal heat is likely due to gravitational contraction as the gas giants cool, contract and release their heat of accretion, but helium condensation is expected to make a significant contribution to the heat budget (see Atreya et al. 2019a for additional details). Gravitational contraction is expected to be negligible for the smaller icy giant planets, yet Neptune's heat excess is similar to or greater than that of the gas giants, with the planet emitting nearly 2-3 times the energy absorbed from the Sun, whereas the internal energy of Uranus is less than 10%. What role might helium condensation play in the heat balance of the icy giant planets is a mystery. But, first, a precise determination of the helium abundance in their atmospheres is required, as was done at Jupiter by the Galileo probe. The presently available values of He mole fraction on Uranus and Neptune from Voyager have very large uncertainties (Figure 8, Table 2). The uncertainty is inherent to the only method that could be applied to Uranus and Neptune – a combination of infrared spectra and the radio refractivity measurements – which depends on (poorly constrained) composition and temperature data (Gautier et al. 1995). Similarly, indirect retrievals of the He/$H_2$ ratio from Voyager at Jupiter and Saturn had large uncertainties also. Only with direct, in situ measurements on Galileo probe, has it been possible to determine the precise He/$H_2$ ratio on Jupiter (Table 2). A precise determination of the helium and neon abundances in the upper tropospheres of Uranus and Neptune is also essential to understand the interior processes and heat balance of these icy giants.

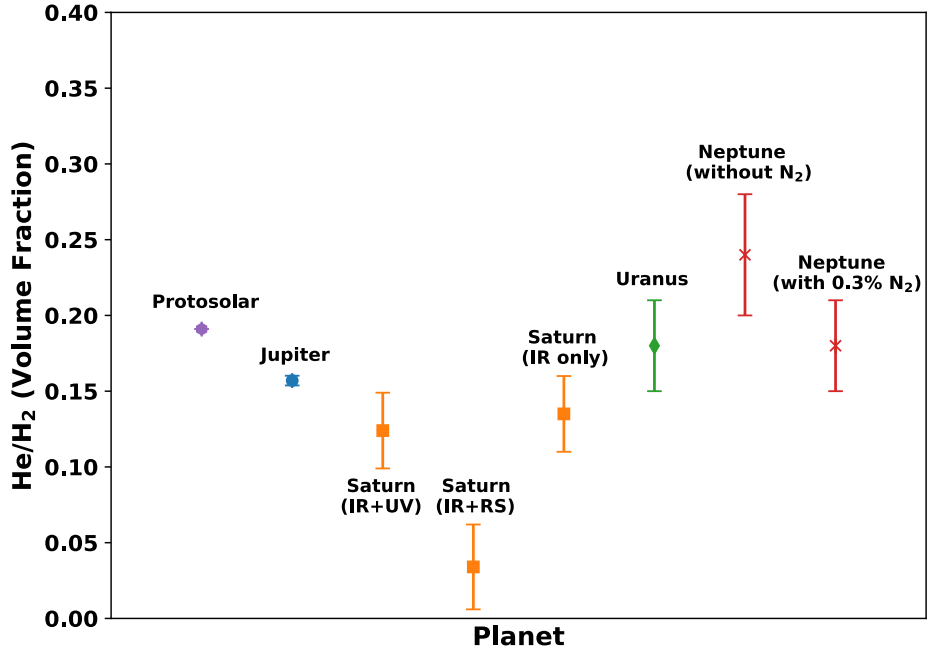

**Figure 8**. He/H$_2$ ratio in the atmospheres of the giant planets and Sun.

Table 2. Helium fraction in the atmospheres of Jupiter, Saturn, Uranus and Neptune

| Object | He Mole Fraction | He/H$_2$ (Volume Fraction) | He/ H$_2$ (Mass Fraction) [g] |
|---|---|---|---|
| Protosolar[a] | — | 0.191 | 0.276 |
| Jupiter[b] | 0.136±0.005 | 0.157±0.003 | 0.238±0.003 |
| Saturn | | | |
| IR+UV[c] | 0.110±0.020 | 0.124±0.025 | 0.199±0.031 |
| IR + RS[d] | 0.033±0.027 | 0.034±0.028 | 0.064±0.055 |
| IR only[e] | 0.118±0.023 | 0.135±0.025 | 0.212±0.031 |
| Uranus[f] | 0.150±0.025 | 0.180±0.030 | 0.265±0.031 |
| Neptune | | | |
| without N$_2$[f] | 0.190±0.030 | 0.240±0.038 | 0.324±0.036 |
| with 0.3% N$_2$[f] | 0.150±0.025 | 0.180±0.030 | 0.265±0.031 |

[a]Asplund et al. (2009); [b]von Zahn et al. (1998); [c]Koskinen and Guerlet (2018); [d]Conrath et al. (1984); [e]Conrath and Gautier (2000); [f]Gautier et al. (1995)
[g]Conversion of volume fraction to mass fraction: $\frac{M_{He}}{M_{H_2}+M_{He}} = \frac{4[He/H_2]}{2[H_2/H_2]+4[He/H_2]}$, where $M_x$ is the molecular mass of the gas and He/H$_2$ is the volume fraction of He to H$_2$.

### 3.3.2 Heavy Noble Gases – Ar, Kr, Xe

How the volatiles were delivered is an important consideration for the formation models. In the following, we briefly summarize possible delivery scenarios, focusing only on the three heavy noble gases, Ar, Kr and Xe, whose abundances depend on the particular delivery method. Additional details are given in Mousis et al. (2020).

*(i) Volatiles Adsorption On Amorphous Ice*
The building blocks of the giant planets may have been agglomerated from pristine materials originating from the interstellar medium. This scenario has been proposed to account for the (relatively) homogeneous volatile enrichments measured by the Galileo probe (Figure 3, Table 1) in Jupiter's atmosphere (Owen et al. 1999) and could arguably apply to any of the four giant planets. In this case, laboratory experiments on amorphous ice show that the fractionation effects are negligible between the adsorption and release of the volatiles at thermodynamic conditions relevant to the PSN (Bar-Nun et al. 2007). This implies the volatile abundance ratios should remain identical once released compared to those in the gaseous mixture prior to adsorption by amorphous ice. Based on these properties, and under the assumption that the volatiles adsorbed on amorphous ice initially formed a protosolar mixture prior to trapping, Ar, Kr, and Xe should show enrichments similar to the one measured for C in the atmospheres of Uranus and Neptune (Figure 9). Interestingly, a similar atmospheric signature can be predicted in the case of the gravitational collapse model. The two planets would have initially formed from the collapse of a cloud made of protosolar gas and the settling of solid grains conjugated with the photoevaporation of the envelope could lead to important volatiles enrichments, with no fractionation mechanism identified so far (Mousis et al. 2018a). However, as discussed earlier, the gravitational instability model faces numerous difficulties.

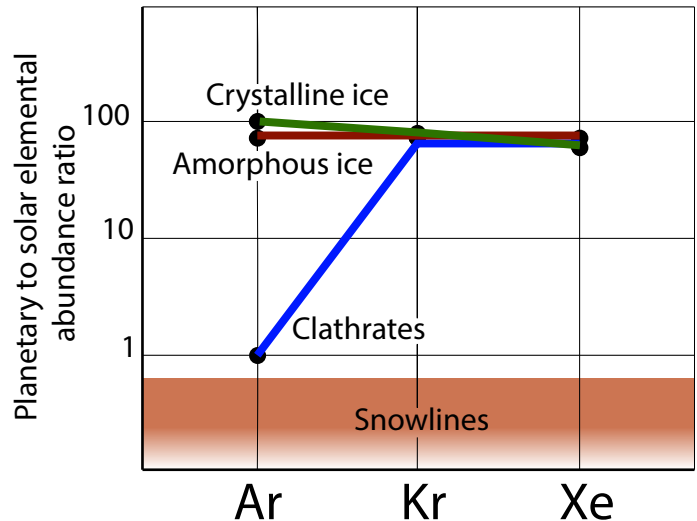

**Figure 9.** Qualitative differences between the enrichments in volatiles predicted in Uranus and Neptune by the different formation scenarios (calibrations based on carbon (C/H) from the $CH_4$ data). The resulting enrichments of the heavy noble gases, Ar, Kr and X, are shown in green (crystalline ice), brown (amorphous ice), blue (clathrates), and shaded brown (snowlines). [Mousis et al. 2020.]

*(ii) Crystalline Ices And Clathrates*
The volatile enrichments in Jupiter (Figure 3, Table 1) have also been interpreted as the result of the giant planet's formation from building blocks agglomerated from (a) clathrates (Gautier et al. 2001; Hersant et al. 2004; Mousis et al. 2009), (b) crystalline ices/pure condensates (Mousis et al. 2009), and also (c) from a mixture of these two phases (Mousis et al. 2012, 2014). The latter hypothesis is supported by the composition of comet 67P/Churyumov-Gerasimenko measured by the ROSINA mass spectrometer aboard the Rosetta spacecraft (Mousis et al. 2016, 2018b). Two extreme members can be considered: agglomeration of Uranus and Neptune's building blocks from clathrates or from pure condensates only. In the former case, an interesting property of the clathration mechanism is the poor propensity of Ar for trapping in the clathrate cages (Lunine and Stevenson 1985; Mousis et al. 2016). In this case, the accretion of clathrate-rich planetesimals or pebbles by Uranus and Neptune should be reflected by a protosolar argon abundance in their envelopes. In the case of building blocks agglomerated from a mixture of various crystalline ices, the condensation sequence of the heavy noble gases indicates that pure crystalline ices of Xe, Kr, and Ar form at ~38 K, 29 K, and 22 K in the PSN, respectively (Mousis et al. 2009). Instead of having a flat enrichment profile for the noble gases in the icy giant atmospheres as for the amorphous case, one should expect Xe/Ar and Kr/Ar ratios about 1.6 and 1.3 greater than their respective protosolar ratios, with a Kr enrichment similar to that of C (see Mousis et al. 2020 for details).

*(iii) Snowlines*

To explain the high C enrichment in Uranus and Neptune (Sec.3.1), as well as their "apparent" N deficiencies (Sec. 3.1) and their presumably high oxygen abundances, it has been proposed that the two planets formed in the vicinity of the CO pure condensate snowline in the PSN (Ali-Dib et al. 2014). By doing so, the amount of $N_2$ ice was negligible in the feeding zone of the two icy giants, contrasting with the large amount of CO-rich solids present. In this mechanism, because the two icy giants were not located at the exact positions of the Ar, Kr, and Xe snowlines, these noble gases should show depletions in their formation regions, also implying subsolar abundances in the envelopes of Uranus and Neptune (Figure 9, and Mousis et al. 2020). Another scenario suggests that Jupiter formed in the vicinity of the amorphous ice snowline, location at which amorphous water crystallizes and releases the adsorbed volatiles to the PSN gas phase. This scenario proposes that Jupiter accreted water in solid form and the other volatiles in gaseous phase. While this mechanism works for Jupiter, it may run into difficulty for explaining the high C enrichment in Uranus and Neptune.

**3.4 Isotopic Ratios**

Isotopic ratios reveal the nature of the source material of the planetary building blocks and the atmospheric evolution. The isotopic ratios of argon, xenon and krypton on Jupiter measured by the Galileo probe confirmed them to be in the solar proportions (Table 3; Mahaffy et al. 2000). The solar composition was affirmed also by the uniform enrichment of all measured heavy elements within the range of uncertainty (global O/H to be determined by Juno). None of the isotopic ratios have yet been measured for the icy giants, as they require entry probes. Precise measurements of the noble gas isotopes will be essential to determine whether the icy giant planets also reflect the solar composition, like Jupiter, or altered by post-formation pollution. The D/H ratio is also important in this regard. The D/H ratio in hydrogen measured by Herschel-PACS and the Infrared Space Observatory (Feuchtgruber et al. 2013) is very similar for Uranus and Neptune (Table 3). It is found to be enriched by about a factor of 2 compared to the Jupiter and protosolar D/H (Table 3), but much lower than the range of D/H in comets (Figure 10), which led the authors to conclude that the interiors of Uranus and Neptune are rock-dominated, with ice comprising a mere 14-32% by mass. The D/H ratio has since been measured in a number of new comets, many with D/H values twice as high as the upper range ($3\times10^{-4}$) assumed previously, which would imply the icy component as low as 7%. However, the actual ice/rock ratio depends on the history of planetary formation and evolution, mixing in the interior, etc., all of which are poorly constrained at this time. Measurements of D/H in $CH_4$, abundances and isotopic ratios of key elements, gravity and magnetic field data, amongst others, are required on any future icy giant planet mission to test various hypotheses and arrive at a realistic ice/rock ratio as well as their solid/gas ratio. Isotopic ratio of C in $CH_4$ would be fairly straightforward to measure with a mass spectrometer, considering virtually no overlap from $NH_3$ or $H_2O$ fragments due their highly depleted abundances. Comparison of $^{13}C/^{12}C$ with other solar system objects is important also to understand the history of formation and evolution of the icy giant planets. $^{15}N/^{14}N$ is also desirable, but unlikely to be measured due to very low abundance of $NH_3$ in the sensible part of the atmosphere.

Table 3. Elemental isotopic ratios in the Sun, Jupiter, Saturn, Uranus and Neptune

| Elements | Sun[a] | Jupiter[b] | Saturn[b] | Uranus[c] | Neptune[c] |
|---|---|---|---|---|---|
| $^{13}C/^{12}C$ | 0.0112 | 0.0108±0.0005 | 0.0109±0.001 | NA | NA |
| $^{15}N/^{14}N$ | 2.27±0.08×10$^{-3}$ | (2.3±0.03)×10$^{-3}$ (0.8–2.8 bar) 1.9(+0.9,-1.0)×10$^{-3}$ (0.2–1.0 bar) | <2.0×10$^{-3}$ (900 cm$^{-1}$ channel) <2.8×10$^{-3}$ (960 cm$^{-1}$ channel) | NA | NA |
| D/H | (2.0±0.5)×10$^{-5}$ [a] (2.1±0.5)×10$^{-5}$ [d] | (2.6±0.7)×10$^{-5}$ [e] (2.25±0.35)×10$^{-5}$ [f] | 1.7(0.75,-0.45)×10$^{-5}$ [f] | (4.4±0.4)×10$^{-5}$ | (4.1±0.4)×10$^{-5}$ |
| $^{36}Ar/^{38}Ar$ | 5.5±0.0 | 5.6±0.25 | NA | NA | NA |
| $^{136}Xe/Xe$ | 0.0795 | 0.076±0.009 | NA | NA | NA |
| $^{134}Xe/Xe$ | 0.0979 | 0.091±0.007 | NA | NA | NA |
| $^{132}Xe/Xe$ | 0.2651 | 0.290±0.020 | NA | NA | NA |
| $^{131}Xe/Xe$ | 0.2169 | 0.203±0.018 | NA | NA | NA |
| $^{130}Xe/Xe$ | 0.0438 | 0.038±0.005 | NA | NA | NA |
| $^{129}Xe/Xe$ | 0.2725 | 0.285±0.021 | NA | NA | NA |
| $^{128}Xe/Xe$ | 0.0220 | 0.018±0.002 | NA | NA | NA |
| $^{20}Ne/^{22}Ne$ | 13.6 | 13±2 | NA | NA | NA |
| $^{3}He/^{4}He$ | 1.66×10$^{-4}$ (1.5±0.3)×10$^{-4}$ (meteoritic) | (1.66±0.05)×10$^{-4}$ | NA | NA | NA |

NA: Not available
[a]Asplund et al. (2009), updated from Rosman and Taylor (1998); [b]Atreya et al. (2019a) and references therein; [c]Feuchtgruber et al. (2013); [d]Geiss and Gloeckler (1998); [e]Mahaffy et al. (1998); [f]Lellouch et al. (2001).

## 4 Measurements

Only entry probes can measure the volatiles containing key heavy elements, helium and the isotopic ratios required for understanding the formation, evolution and migration scenarios of the giant planets. Elemental abundances of Jupiter are well covered between Galileo and Juno. Only C/H has been measured so far on Uranus and Neptune, and even that should be treated with caution, besides having large uncertainty (Sec. 3.1). Many more types of volatiles condense on the colder icy giant planets than Jupiter (Sec. 3.1 and 3.2) and their well-mixed abundances are much deeper, so a different approach is necessary.

**4.1 Insights Concerning Bulk Abundances of Water, Ammonia and Hydrogen Sulfide (O, N, S)**

The discussion of bulk composition in Sec.3 shows that the condensible volatiles, $NH_3$ and $H_2S$, are greatly depleted down to at least several tens of bar levels in the atmospheres of Uranus and Neptune. Water could also be depleted due to sequestration in purported liquid water ocean or an ionic/superionic ocean. Even in the absence of such oceans, the well-mixed region of water would lie at kilobar levels. If the O/H is as enriched as C/H, i.e. 80±20× solar, the water cloud at Uranus would form at ~1 kilobar level based on equilibrium cloud condensation model (Figure 11; Neptune is very similar). However, well-mixed water may actually lie much deeper, perhaps even at tens of kilobar level, as Juno found for ammonia in Jupiter's atmosphere, where the well-mixed $NH_3$ lies several tens of bars below its condensation level of ~0.7 bar except near the equator (Bolton et al. 2017, Li et al. 2017). This means that with the exception of perhaps methane, direct in situ measurement of the bulk abundances of condensible volatiles, $NH_3$, $H_2S$ and $H_2O$, is impractical, as current technology limits entry probe measurements to 5-10 bars at Uranus and Neptune.

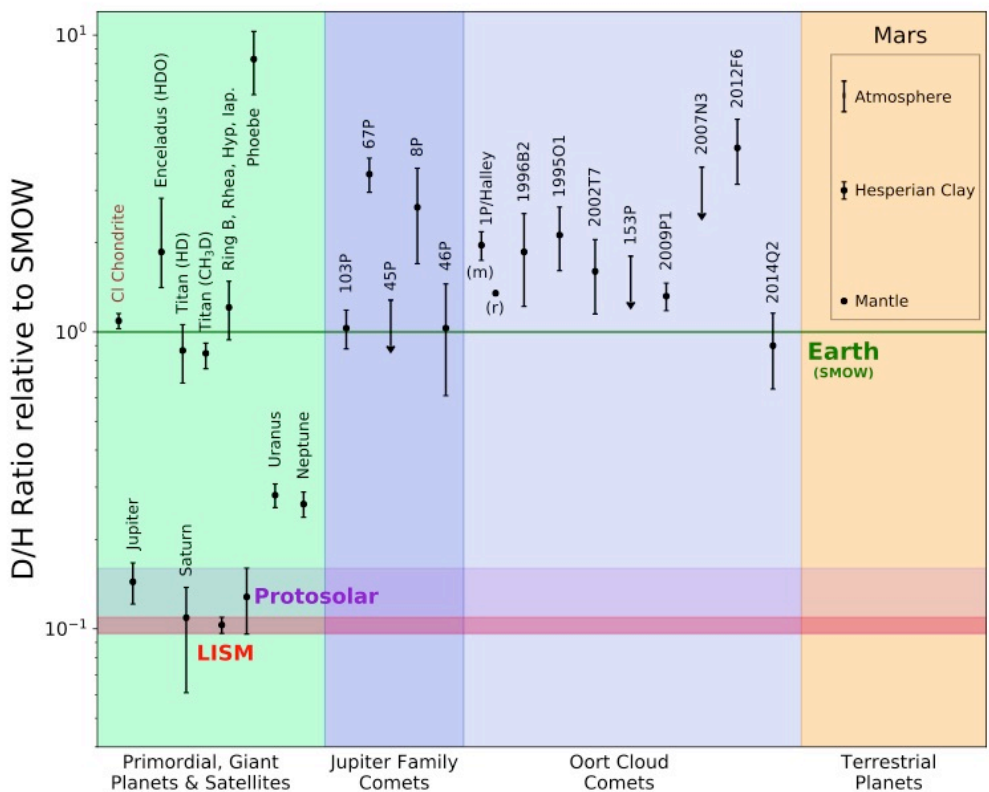

**Figure 10**. Deuterium to hydrogen ratio relative to Earth's Standard Mean Ocean Water (SMOW) value ($1.558 \times 10^{-4}$). Saturn's Ring B, Rhea, Hyperion and Iapetus have similar values, all close to SMOW, so the plot shows an average of those values. Comet names have been abbreviated in the plot, and they are 103P: Hartley 2, 67P: Churyumov–Gerasimenko, 45P: Honda-Mrkos-Pajdusakova, 8P: Tuttle, 46P: Wirtanen, 1P: Halley (m) measured; (r) reinterpreted, 1996B2: Hyakutake, 1995O1: Hale-Bopp, 2002T7: LINEAR, 153P: Ikeya-Zhang, 2009P1: Garrad, 2007N3: Lulin, 2012F6: Lemmon, 2014Q2: Lovejoy. Venus D/H is off the scale. Its values are: $(5-6.5) \times 10^{-2}$ below the clouds, and $(2.5 \pm 0.5) \times 10^{-2}$ in 70–90 km range, and its ratios to SMOW are $160 \pm 32$ below the clouds, and 320–417 in 70-90 km range.

Even without direct measurements of N, S and O elemental ratios, however, valuable insights would be gained into chemical processes operating within the exotic interiors of the icy giants. Measuring $H_2S$ (and perhaps even $NH_3$) mixing ratios in the p < 10 bar region would provide quantitative constraints on the relative affinities of these species to deeper water layers, and may perhaps distinguish between $NH_4SH$ or $(NH_4)_2S$ condensation at the probe descent path. A single probe can sample only one location, but measurement of volatiles in a second location, perhaps by a miniaturized entry probe (Sayanagi et al. 2018), would help to distinguish between chemical and dynamical atmospheric effects.

Spatial variation in volatile concentrations is also very well constrained by ground-based or orbiter observations, but remote sensing measurements are often challenged by degeneracies between temperature and compositional variation in the atmosphere (e.g., Fletcher et al. 2016). Probe measurements are therefore extremely valuable as a ground truth capable of breaking such degeneracies, as evidenced by the large number of Jupiter remote sensing studies referencing probe measurements of He, $CH_4$, and thermal structure (von Zahn et al. 1998, Niemann et al. 1998, Wong et al. 2004, Seiff et al. 1998).

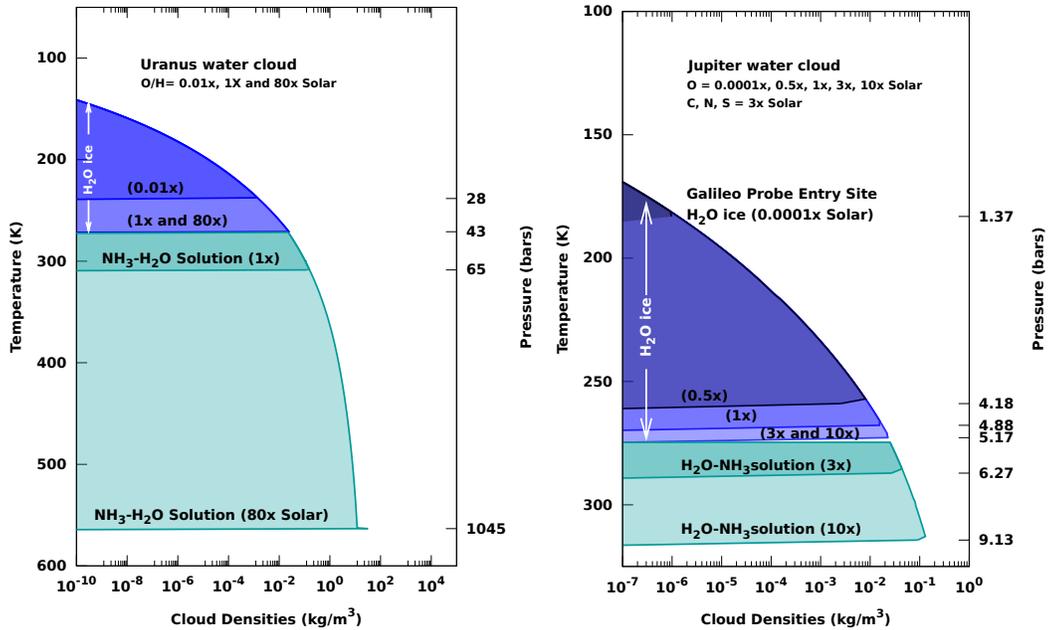

**Figure 11**. Comparison of the depth of a water cloud on Uranus and Jupiter for different abundances of well-mixed water (O/H) relative to its solar value. On Jupiter, the water cloud would form at ~6 bar level if O/H is similar to the other elements, i.e. ~3× solar (Table 1), whereas on Uranus it would form at a kilobar level is O/H = C/H, which is ~80× solar (Table 1). See Fig. 5 caption for discussion of cloud densities and cloud base pressure levels.

### 4.2 Noble Gases are Key for The Icy Giant Planets

The heavy noble gases, Ar, Kr and Xe − being chemically inert and non condensible on the icy giants − are expected to be well-mixed everywhere below the homopause, not subject to any dynamical effects unlike the condensible volatiles (see Sec. 3.3.1 for He and Ne). However, their abundances are low, with solar Xe/H at only a tenth of a ppbv and Kr/H just ten times higher. Thus, measurements done at atmospheric pressures of a few bars would give greater signal to noise ratio, allowing accurate determination of all noble gases and their isotopes, as on the Galileo probe at Jupiter. Another set of measurements a few bars below would be important for confirmation and robustness of the data. Unlike the condensible species, noble gas measurements done in a single location of an entry probe should be representative of their global values.

### 4.3 Complementary Observations on Composition and Atmospheric Dynamics

#### 4.3.1 Oxygen Elemental Abundance from CO?

As seen earlier in this section, water is too deep for in situ investigation, hence unsuitable for determining the oxygen elemental abundance of the icy giant planets. CO is another reservoir of oxygen, albeit a minor one, in the atmosphere of the giant planets. CO is a disequilibrium species in the atmospheres of the giant planets. Its presence in the upper troposphere implies strong convective upwelling from its thermochemical equilibrium level in the deep atmosphere, as is implied also by other disequilibrium species, such as $PH_3$, $GeH_4$ and $AsH_3$, all of which have been detected in the upper troposphere/lower stratosphere of Jupiter and Saturn. The first detection of CO in the atmosphere of Neptune was made at submillimeter wavelength of 345.8 GHz with JCMT (Marten et al. 1993). It was found to be present both in the stratosphere and the troposphere, and the data were consistent with a uniform mixing ratio of $1.2 \times 10^{-6}$. More recent observations derive a stepped vertical profile, with a high upper atmospheric CO concentration and a smaller concentration in the deeper atmosphere. These stepped profiles, when applied to recent millimeter and submillimeter Neptune spectra, are consistent with deep CO concentrations of 0.1 to 0.6 ppm (Lellouch et al. 2005, Hesman et al. 2007, Moreno et al. 2011, Luszcz-Cook and de Pater 2013, Teanby et al. 2019). However, the results are model dependent. Luszcz-Cook and de Pater (2013) found that their observations could be fit with zero tropospheric CO if there was additional absorption from 10× solar or more $H_2S$ and the apparent absence of $NH_3$ absorption in the microwave spectrum (Sec. 3.1). Furthermore, Teanby et al. (2019) found that stepped profiles may overestimate the deeper tropospheric CO concentration, because their results only required nonzero CO in the upper part of the troposphere at P < 0.5 bar. Herschel/SPIRE observations (15–52 $cm^{-1}$) set the most stringent upper limit of 2.1 ppb in the upper troposphere of Uranus (Teanby and Irwin 2013).

Models of disequilibrium species concentrations often combine complex thermochemical kinetic schemes with parameterized one-dimensional (vertical) diffusion. The models span the vertical range between the observable troposphere down to levels where thermochemical equilibrium describes the partitioning between different molecular species, such as CO and $CH_4$ for carbon. One goal of this modeling approach is to relate CO observations to the bulk C/H and O/H ratios. On Jupiter, Wang et al. (2016) combined the Bézard et al. (2002) tropospheric CO volume mixing ratio of 1 ppb with a deep eddy mixing parameter Kzz near $10^8$ $cm^2$ $s^{-1}$. They constrained the O/H ratio to the 0.1–0.75× solar range using the reaction network of Visscher and Moses (2011), or the 3–11× solar range using reaction network from Venot et al. (2012). Spectroscopic constraints on the water cloud condensation level in Bjoraker et al. (2018) preclude the subsolar O/H derived with the Visscher and Moses (2011) reaction network.

On Uranus, the CO volume mixing ratio (VMR) upper limits have not been able to set useful constraints on O/H (Cavalié et al. 2014, Teanby and Irwin 2013). On Neptune, a CO VMR of 1 ppm was used by Lodders and Fegley (1994) to set limits of O/H = 670× solar (updated to the Asplund et al. 2009 values). Luszcz-Cook and de Pater (2013) used updated CO measurements, rate-limiting reactions, and effective mixing length to revise the O/H ratio downward to 500× solar, but when they included $H_2S$ absorption, their model was consistent with zero tropospheric CO, similar to the preferred model of Teanby et al. (2019). A more comprehensive discussion may be found in Cavalié et al. (2017, and this issue).

Above discussion illustrates the difficulty of deriving the O/H ratio from CO. The results are highly model dependent and results are inconclusive due to lack of meaningful constraints on input parameters such vertical mixing, thermochemistry, reaction kinetics, etc. Further complications could arise depending on the relative positions of the ionic ocean and the quenching level for $H_2O$/CO thermochemical equilibrium. Nevertheless, modeling efforts should continue considering that CO may be the only candidate for getting at possible range of oxygen elemental abundance on the icy giant planets. A gas chromatograph together with mass spectrometer (GCMS) has a good chance of making the CO measurement in situ in the troposphere in order to discriminate between $N_2$ and CO both at 28 daltons. Remote sensing observations of the other disequilibrium species from the orbiter can provide some insight into diffusion from their thermochemical equilibrium levels in the deep atmosphere to the lower stratosphere.

**4.3.2 Microwave Remote Sensing: Dynamics from Vertical Profiles of $H_2O$ and $NH_3$**

Juno's microwave radiometer (MWR) is designed to determine the bulk abundance of water in the deep atmosphere of Jupiter (Janssen et al. 2017), hence the oxygen elemental ratio. The contribution function of MWR's longest wavelength channel (50 cm, 0.6 GHz) peaks at ~100 bars, with half the contribution still coming from as deep as 1 kilobar. A similar instrument on an orbiting spacecraft could be a useful tool for measuring both water and the ammonia abundances down to deep atmospheric levels of Uranus and Neptune. An advantage of the Juno MWR over ground-based observations is that it can conduct atmospheric observations inside the radiation belts, which produce non-thermal (synchrotron) emission. Although synchrotron radiation is not a concern for ground based observations of the icy giants (Fig. 6), spacecraft observations will be able to achieve higher spatial resolution. Ground-based millimeter/microwave observations show hints of spatial variation to levels as deep as 30-50 bar, but spatial resolution is limited at the longest (and deepest-probing) wavelengths where the planet is only a few times larger than the beam size (de Pater et al. 2014, Tollefson et al. 2019b). But measurements reaching down to kilobar levels may not be deep enough for the well-mixed water, $NH_3$ and $H_2S$, as discussed above. Thus even microwave remote sensing is unlikely to yield the O, N or S elemental ratios. On the other hand, a global map of vertical distribution of $H_2O$, $H_2S$ and $NH_3$ measured by a microwave radiometer on an orbiter can provide exceptional data on the dynamical processes down to kilobar levels. Interpretation of this type of orbiter data would be greatly facilitated by probe measurements of temperature and composition (particularly microwave-absorbing species) in the upper part of the troposphere. A detailed study of the feasibility and science return of microwave radiometry on an orbiter at Uranus and Neptune is required.

## 5 Implementation: Possible Mission Design and Architecture Concepts

The focus of this paper is on fundamental questions such as the formation and migration of the icy giant planets, which only entry probes can address. However, in situ measurements must be placed in the broader context of Uranus/Neptune system science, which require complementary remote sensing observations from an orbiter. Ideally, such a mission would be similar to NASA's Galileo orbiter-probe mission at Jupiter. Less costly missions than a full-up Flagship class mission, such as a Juno-like New Frontiers class mission but with an entry probe, may be feasible to address limited but focused science objectives. A good insight into possible missions to the icy giants can be had by looking at the main conclusions of NASA's pre-Decadal Ice Giants Science Definition Team study (IG SDT Report) for which NASA's Jet Propulsion Laboratory (JPL) developed an array of mission designs and system architectures. The SDT-recommended minimal concept from that study is an orbiter with probe to either Uranus or Neptune. The most comprehensive concept studied is a dual-spacecraft, dual-planet mission that could explore both Uranus and Neptune. This section will provide an overview of a typical stand-alone Neptune mission because it includes all elements of these three architectures. This will be followed by a brief description of the differences for a typical Uranus stand-alone and dual-spacecraft, dual-planet concepts. Launch and technology opportunities are also summarized.

**5.1 A Typical Stand-alone Neptune Mission**

Neptune's distance from Earth and the interplanetary travel time to get there are significant architectural drivers given the lifetime constraints of NASA's eMMRTG (enhanced Multi-Mission Radioisotope Thermoelectric Generator) power system. Neptune can be further than Pluto, depending upon where Pluto is in its orbit. Early estimates indicate that a Neptune orbiter with the payload complement specified by the SDT would exceed 1,000 kg dry mass. This mass, coupled with mass of a Neptune atmospheric probe, leads to the conclusion that a solar electric propulsion or SEP-based architecture is required to deliver the spacecraft into orbit and complete the science mission within the design life of the radioisotope power system. This cannot be achieved with a chemical propulsion alternative. Figure 12 illustrates key elements of a typical Neptune flight system architecture. Payload elements are notional and not meant to imply endorsement or flight readiness.

The mission design shown in Figure 13 can be divided into three mission phases: interplanetary cruise, probe release and data relay, and Neptune orbit insertion.

The interplanetary trajectory relies on a 25-kW SEP stage powered by 3 NASA Evolutionary Xenon Thrusters (NEXT ion engines) to propel the spacecraft through the inner solar system (see IG SDT Report for additional details). No additional maneuvers are required between probe release and NOI. Post-launch, the spacecraft applies the SEP stage to gain momentum, perform an Earth flyby followed by a Jupiter flyby, and transition onto a long coast phase toward Neptune. The SEP stage makes up for the relatively low launch energy through use of high propellant efficiency and continuous thrust arcs.

At a range of ~6 AU, where the solar insolation becomes insufficient for SEP to be effective, the stage is jettisoned. Releasing it before Neptune Orbit Insertion (NOI) reduces mass and enables significant propellant savings.

The probe is released ~60 days prior to Neptune atmospheric entry. A probe-targeting maneuver (PTM) is performed prior to probe release, followed by an Orbiter Divert and Periapsis Targeting Maneuver to achieve the desired conditions for orbit insertion. The probe enters Neptune's atmosphere at an entry flight path angle (EFPA) of -20 degrees. While this relatively shallow EFPA constrains the probe-orbiter telecommunication geometry, it is required to reduce deceleration and heat loads. Due to the relatively low data rate requirements, this geometry is more than adequate for full mission data return. The probe descent to 10 bar lasts for ~1 hr, of which the first ~30 mins represent the entry sequence.

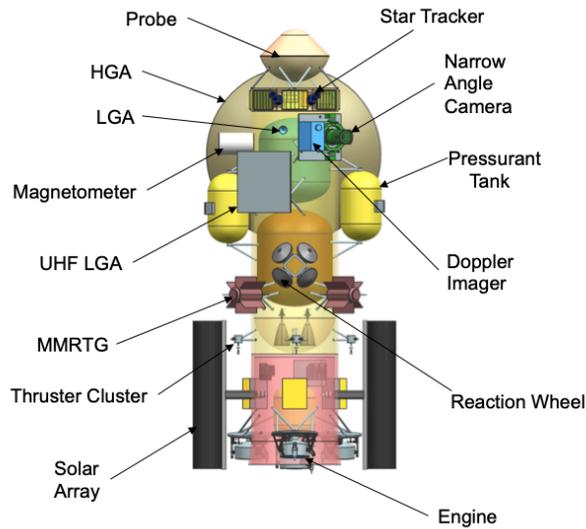

**Figure** 12. Conceptual design of eMMRTG-powered Neptune orbiter with SEP and atmospheric probe. Payload is notional.

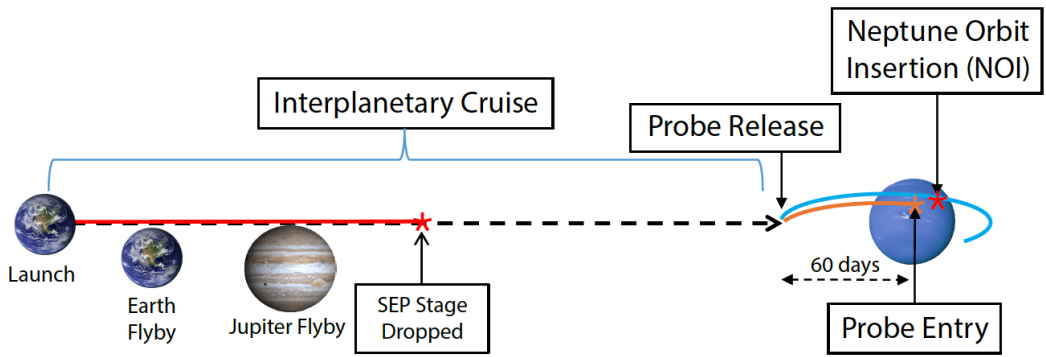

**Figure** 13. Possible scenario for a typical Neptune mission: Solar electric propulsion (SEP) architecture.

Following probe relay, the orbiter performs a large orbit insertion maneuver (~2.7 km/s delta v for Neptune, ~2.0 km/s for Uranus) at an altitude of ~1.05 Neptune Radii and enters in a 252-day retrograde orbit for a typical Neptune mission studied. The orbit insertion altitude is chosen to mitigate potential ring crossing issues and to lower the NOI delta v. Figure 14 shows the NOI location, orbiter path, probe approach and probe descent trajectory.

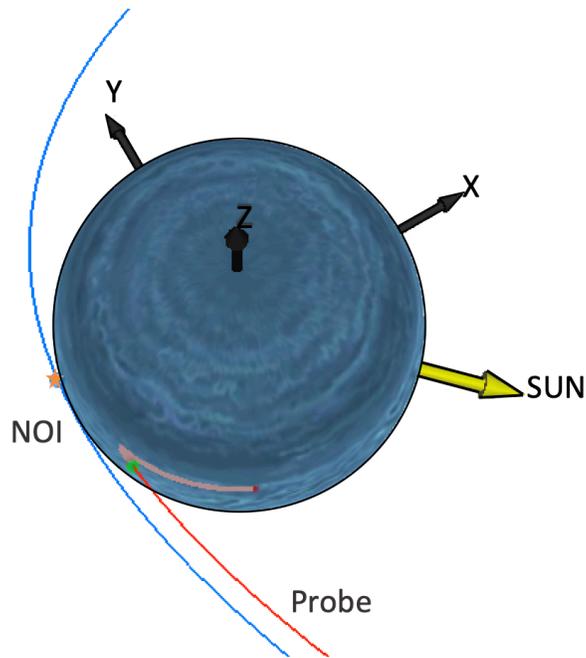

**Figure 14**. Neptune orbit insertion accommodates probe data relay during descent

The atmospheric probe design draws on heritage from Galileo and Pioneer Venus probes, using current state of the art technologies and instrument designs. The probe is spin stabilized during its coast to the planet and is powered by primary batteries. Survival heating after probe release and during the 60-day coast period is provided by radioisotope heater units (RHUs). The overall configuration, including SDT example instrument complement is illustrated in Figure 15.

The probe descent module is a truncated sphere, approximately 73 cm in diameter. The descent module is vented, allowing an equalization of pressure inside the probe with the external atmosphere during its descent. Apertures in the probe provide instrument access to the atmosphere. Telecom uses a flat patch antenna on the top of the probe to maintain a communications link with the orbiter during the probes ~60-minute science retrieval.

The probe entry system consists of a Galileo 45-degree sphere-cone heat shield scaled to 1.2 m in diameter, and a spherical backshell with a radius of curvature originating at the vehicle center of gravity (CG). The heat shield uses a 3D weave of blended carbon/phenolic yarns (Heat Shield for Extreme Entry Environment, HEEET) under development by NASA's Ames Research Center. The backshell TPS (Thermal Protection System) is flight proven PICA (Phenolic Impregnated Carbon Ablator). The mass of the probe at entry is estimated to be ~308 kg.

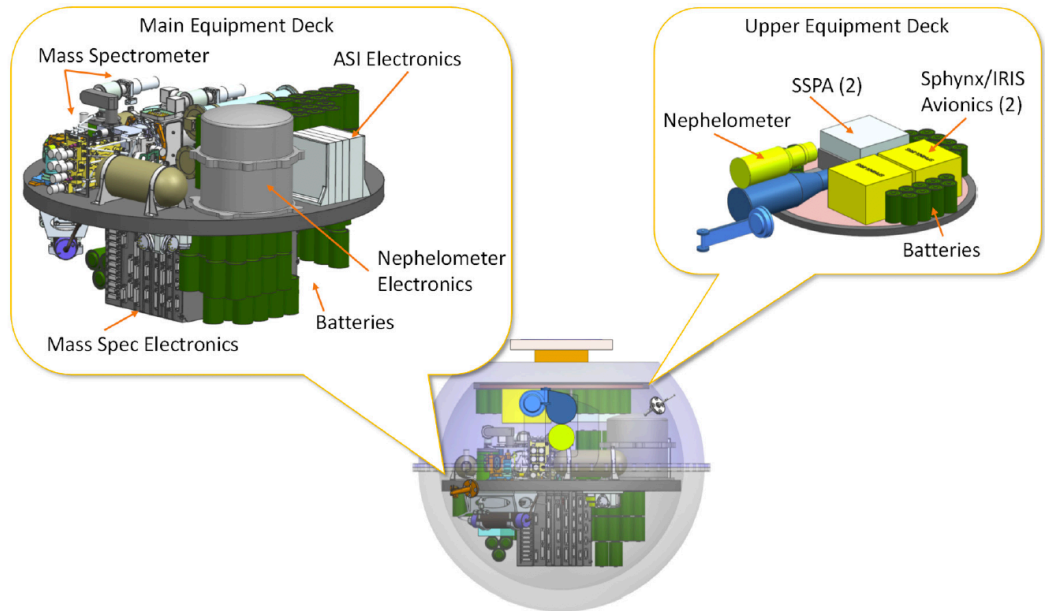

**Figure 15**. Conceptual design of probe descent module is based on mature flight proven elements.

The probe instrument suite as shown in Figure 15 comprises four notional instruments:

- Mass Spectrometer (MS or GCMS)
- Atmospheric Structure Instrument (ASI)
- Nephelometer
- Ortho/Para Hydrogen Experiment (OPH)

Only two instruments on the probe – MS/GCMS, and ASI – are absolutely critical to address the fundamental questions of planetary formation. A helium detector may be included for redundancy, as on the Galileo probe at Jupiter, but is not required. MS/GCMS and ASI have a long heritage, having flown successfully on a number of planetary spacecraft including Galileo, Huygens, Pioneer Venus, MSL, etc. Other instruments may also be considered for inclusion on payload depending on available resources and compelling science drivers.

While the baseline launch vehicle assumption for the scenario presented in this paper is a Delta IV-Heavy, the availability of Space Launch System (SLS) or another rocket with similar capability would allow:

(a) Reduced flight times and/or increased delivered mass to either icy giant planet. This allows additional tradeoffs between cost and science return.
(b) Two-planet, two-spacecraft missions on a single launch vehicle.

**5.2 A Typical Stand-alone Uranus Mission**
A Uranus concept would be similar to that discussed for Neptune however the solar electric propulsion (SEP) stage on the carrier spacecraft would not be necessary because Uranus is significantly closer to Earth thereby reducing flight times enough to meet the NASA eMMRTG design lifetime constraint. The basic spacecraft architecture, instrument complement and atmospheric probe would be the same.

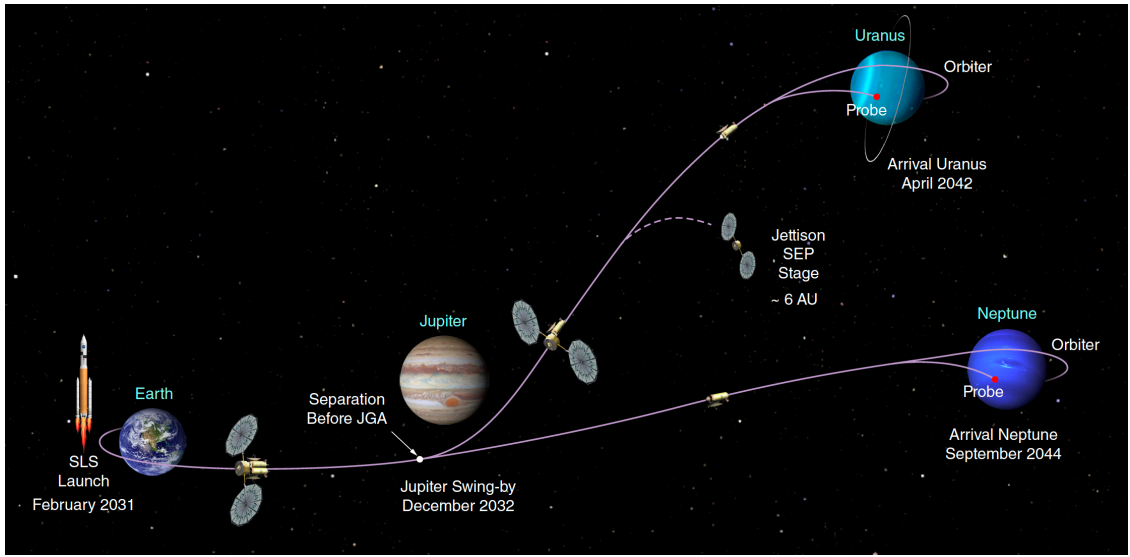

**Figure 16.** Dual-spacecraft, dual-planet mission scenario.

**5.3 Dual Planet Mission**

While there are multiple options that can be considered for a dual planet mission, the example scenario studied is shown in Figure 16. Both spacecraft could launch together on a NASA Space Launch System or another rocket of equivalent capability. Both spacecraft are stacked on top of the SEP stage and remain that way throughout the inner solar system tour. Just before arriving in the vicinity of Jupiter, they separate and each does its own Jupiter gravity assist (JGA) to put them on their respective trajectories to Uranus and Neptune. The SEP stage stays with the Uranus spacecraft for additional post JGA thrusting and then jettisoned at ~6 AU beyond which SEP stage is no longer effective. While both spacecraft can be launched on the SLS, there is no scientific penalty to launching two-planet missions on different launch vehicles several years apart.

**5.4 Launch Opportunities**

Launch to an icy giant planet is generally possible any year, but there are significant variations in performance (results from more than 10,000 trajectories can be found in IG SDT Report). As shown in Table 4, the availability of Jupiter gravity assist maximizes delivered mass resulting in preferential launch windows for Uranus missions in the 2030–2034 timeframe and a corresponding window of 2029– 2030 for Neptune. In these favorable periods chemical trajectories could deliver ample mass for the Uranus missions studied in an 11-year flight time, using a launch performance capability similar to the Atlas V 551. Neptune trajectories utilizing solar electric propulsion (SEP) can deliver a similar mass to Neptune orbit in 13 years using launch performance capability similar to the Delta IV-H. There are no all-chemical trajectories to Neptune, even using a Delta IV-H, that yield a mission duration less than 15 years, a design target chosen to be consistent with Radioisotope Power System (RPS) design life and mission reliability. Significant science can be done during gravity assists at a gas giant. If a Saturn flyby is preferred over the Jupiter gravity assist, only trajectories to Uranus are available in the next decade, and launch must occur before mid-2028. The use of SEP for inner solar system thrusting has the potential to significantly reduce flight times to Uranus and/or increase delivered mass. The SEP stage would be used as far out as 6 AU, at which point solar power is insufficient to provide additional thrusting and the SEP stage would be jettisoned. SEP-enhanced mission concept designs also see a slight preference in launch dates corresponding to availability of Jupiter gravity assists, but well-performing trajectories are possible in any year of the period studied.

Table 4. Orbiter-Class Mission: Launch-no-later-than Dates[a]

| | | Uranus | Neptune | Dual |
|---|---|---|---|---|
| Best Launch Years (JGA) | | 2030-2034 | 2029-2030 | 2031 |
| Launch Class | S/C Propulsion | | | |
| Delta IV-H | Chemical | 2035 | 2029 | n/a |
| | SEP | Any | 2030 | n/a |
| SLS-Block1B | Chemical | Any | 2031 | n/a |
| | SEP | Any | Any | 2031 |
| Delta IV-H | SEP | | Any | n/a |
| *Color code[b] | | TOF < 12 yrs | TOF < 13 yrs | |
| (TOF is interplanetary portion only) | | | TOF < 20-25 yrs | |

*Table 4 Footnotes*
[a]
-Best Launch Years (JGA) in row 2 show the "best" years to launch, when most mass margin is possible regardless of launch vehicle or propulsion type, but subject to color-coded TOF limits shown in the bottom two rows. The second and subsequent lines of launch years (rows 4-8) show the "Launch-no-Later-than" years for corresponding LV+PropulsionSystem combination indicated on each row. Launching later than those years would not allow flying an orbiter class mission. Technically, with SEP and Delta IV-H, an orbiter class 12-yr TOF mission to Uranus could be launched in any launch year (row 5); however, the "best" launch years with highest mass margin are 2030-2034 (row 2). Similarly, SEP and Delta IV-H combination allows launching to Neptune in any launch year (row 8), but the TOF would be prohibitively long, 20-25 years.
-JGAs confer a significant advantage in delivered mass and flight time.
-JGA is available for a few years around 2030; next Launch opportunity w/ JGAs is around 2046.
-Launching after indicated dates results in insufficient mass delivered for orbiter-class mission.
[b] Color code: salmon, yellow and brown colors indicate interplanetary portion of time of flight (TOF) cases in table cells.

There are no trajectories that allow a single spacecraft to encounter both Uranus and Neptune. As discussed above, a single SLS launch vehicle could, however, launch two spacecraft, one to each icy giant planet.

**5.5 Technology Considerations**

The IG SDT identified only two technologies as enabling for a mission to either Uranus or Neptune: appropriate thermal protection system for the probe and power source for the spacecraft. Heatshield for Extreme Entry Environment Technology (HEEET) would be enabling for the entry conditions of probes at both Uranus and Neptune. The development of HEEET is complete, and has achieved Technical Readiness Level 6 (Gage et al. 2019; Venkatapathy et al. 2020). The eMMRTG would provide a significant improvement in specific power over the existing MMRTG technology at beginning of life but, more importantly, a much larger gain at end of life, which is critical, given the duration of an icy giant planet mission. Its development is proceeding well (Matthes et al. 2019).

A number of new technologies, while not necessary to enable the mission concepts described, could have an impact on the performance and/or cost of the mission if available: 1) Aerocapture could enable trip times to be shortened, delivered mass to be increased or both. 2) Cryogenic propulsion could have similar but not as pronounced effects. 3) Next generation RPS technologies, with even better specific power than the proposed eMMRTG, such as a segmented modular RTG (aka Next-Generation RTG), could enable more mass or power for instruments or both. 4) Advanced telecommunications technologies/capabilities could dramatically increase the data return from an outer planet mission and 5) Mission operations autonomy technologies could drive down cost and permit more adaptive missions operations than are envisaged in the missions reported here.

# 6 Summary and Conclusions

Only entry probes are capable of determining the abundances and isotopic ratios of elements required to constrain models of the formation of Uranus and Neptune. Noble gases and their isotopic ratios, in particular, are key, and accessible at relatively shallow depths. Noble gas measurements can be carried out at pressures of about 1 bar, but probes deployed to 5-10 bar level would allow sampling at multiple depths for confirmation, as well as collection of data on other volatiles. Well-mixed ammonia, hydrogen sulfide, and possibly water, are likely too deep for any in situ or remote sensing observations. Their lack is not detrimental, however, especially in view of robust data on the noble gases and comparative planetology with the gas giants, especially Jupiter with its full suite of elemental and isotope abundances. The most critical payload elements of an entry probe are a mass spectrometer and an atmospheric structure instrument. Observations of the height profiles of $H_2O$, $NH_3$ and $H_2S$ to the deepest levels possible, e.g. with microwave remote sensing from orbit and Earth, would be valuable to an understanding of the atmospheric dynamics. Measurements of gravity, magnetic field and stratospheric composition and structure from an orbiter are highly complementary to the probe data, and together they would result in robust models of planetary interior and formation. Exploration of Uranus or Neptune requires no new mission enabling technology efforts; development of certain technologies may result in shorter cruise, increased payload mass and more flexible launch opportunities, however. International partnership is most desirable for maximizing the science return of an icy giant planet mission. Possible contribution to a NASA-led mission may include a probe, a second spacecraft for the dual-spacecraft, dual-planet option, cubesats or smallsats for certain specific, targeted measurements. Uranus and Neptune remain as the last major pieces of the outer solar system formation puzzle. Their detailed exploration with an orbiter and probe is thus essential for a comprehensive understanding of the origin and evolution of the solar system, and, by extension, the most common type of planets confirmed to date in the extrasolar systems.

**Acknowledgements**
The work of MDH and KR was performed at the Jet Propulsion Laboratory, California Institute of Technology, under a contract with the National Aeronautics and Space Administration. OM acknowledges support from CNES. We thank John Elliott and Anastassios Petropoulos of JPL for their work on mission design studies.